\documentstyle[twoside,fleqn,espcrc2,epsf,epsfig,here]{article}
 
\newcommand{\AmS}{{\protect\the\textfont2
  A\kern-.1667em\lower.5ex\hbox{M}\kern-.125emS}}
\def\beq{\begin{equation}}
\def\eeq{\end{equation}}
\def\bea{\begin{eqnarray}}
\def\eea{\end{eqnarray}}
\def\gappeq{\mathrel{\rlap {\raise.5ex\hbox{$>$}}
{\lower.5ex\hbox{$\sim$}}}}

\def\lappeq{\mathrel{\rlap{\raise.5ex\hbox{$<$}}
{\lower.5ex\hbox{$\sim$}}}}
\def\PL{ Phys.Lett., }
\def\PRL{ Phys.Rev.Lett., }
\def\NP{ Nucl.Phys., }
\def\PR{ Phys.Rev., }
\def\AJ{ Astrophys.J., }

\title{\bf Learning Physics from the Cosmic Microwave Background}
 
\author{J. Ellis\address{
Theoretical Physics Division,  CERN, CH-1211 Geneva 23}}%

\begin{document}
 
\begin{abstract}
\begin{center}
CERN-TH/98-401~~~~~~~~~~~~~~astro-ph/9902242
\end{center}
The Cosmic Microwave Background (CMB) provides a precious window on
fundamental physics at very high energy scales, possibly including
quantum gravity, GUTs and supersymmetry. The CMB has already enabled
defect-based rivals to inflation to be discarded, and will be able to
falsify many inflationary models. In combination with other cosmological
observations, including those of high-redshift supernovae and large-scale
structure, the CMB is on the way to providing a detailed budget for the
density of the Universe, to be compared with particle-physics
calculations for neutrinos and cold dark matter. Thus CMB measurements
complement experiments with the LHC and long-baseline neutrino beams.

\end{abstract}
 
\maketitle
 
\section{Why the CMB Might be a Good Physics Teacher}

Measurements of the CMB by experiments before COBE, by COBE itself, and by
subsequent experiments, have already amassed an impressive amount of
data~\cite{Smoot}, and
this is set to grow dramatically with future experiments culminating in the MAP
and Planck Surveyer satellites. We already know that the spectrum is very close
to black-body, which imposes important constraints on entropy deposition, late
particle decays, reionization, etc.~\cite{PDG}. We also know that the CMB
is highly
isotropic, providing the best evidence for the relevance of
Friedman-Robertson-Walker (FRW) cosmological models~\cite{FRW}. This
isotropy immediately
raises the horizon problem: why is the Universe apparently so homogeneous and
isotropic on large scales? It is worth recalling that the scale size of the
Universe at the epoch of last scattering is about two orders of magnitude
larger than the horizon size $a_H = 2t$ at that epoch $t$, which is the largest
distance over which a message could have travelled in a conventional FRW
cosmology. So how were the opposite sides of the Universe able to coordinate so
precisely? Small anisotropies in the CMB have been seen: the first to be
discovered was the dipole anisotropy of about $10^{-3}$, which is
conventionally interpreted as a D\"oppler effect due to the velocity of the
Earth in the Machian reference frame provided by the CMB~\cite{dipole}.
More recently, COBE
and its successors have detected the higher-order anisotropies shown in
Fig. 1~\cite{anisotropies},
which promise to teach us a lot of fundamental physics.

\begin{figure}[H]
 \epsfig{figure=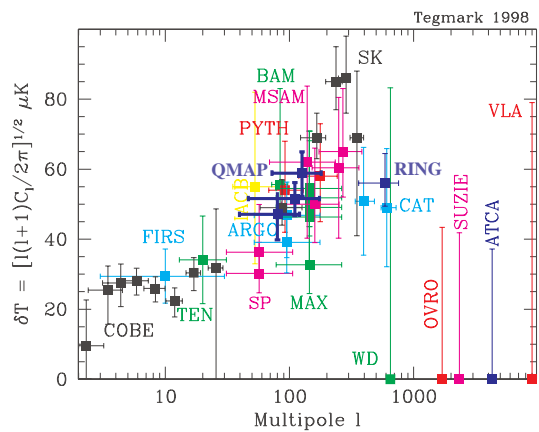,width=7cm}
 \caption[]{{\it Compilation~\cite{anisotropies} of CMB anisotropy
measurements.}}
\end{figure}

These anisotropies are usually interpreted in the context of cosmological
inflation~\cite{inflation}, according to which, at some very early epoch,
the energy density
$\rho$ of the Universe may have been dominated by an (almost) constant term $V$:
\beq
\left({\dot a \over a}\right)^2 = {8\pi G_N \over 3}~\rho ~~-~~ {k\over a^2} ~:~
\rho \simeq V
\label{one}
\eeq
It is easy to see that, if this constant term were dominant, it would generate
an epoch of exponential growth of the scale factor $a$:
\beq
a \simeq a_I~~\exp (H~(t-t_I))~~:~~H = \sqrt{{8\pi G_N \over 3}~V}
\label{two}
\eeq
where $a\simeq a_I$ at the initial time $t_I$ marking the onset of inflation.
If so, the horizon size $a_H$ of the Universe would also have expanded
exponentially, and the entire observable Universe would have been born within
the pre-inflationary horizon:
\beq
a = a_I~e^{H(t-t_I)} \gg  2t~~{\rm even~if}~~ a_I \ll 2t_I
\label{three}
\eeq
During this epoch of exponential expansion, the (approximate)
homogeneity and isotropy of the observable Universe could have been imprinted.
Moreover, the curvature term $-k/a^2$ in (\ref{one}) would rapidly have become
negligible, so that the Universe would become almost critical:  $\Omega \equiv
\rho / \rho_c \simeq 1$, where $\rho_c \equiv \sqrt{3H/8\pi G_N}$. Furthermore,
unwanted particles from the very early Universe, such as GUT
monopoles~\cite{monopoles}, would
have been inflated away beyond the last-scattering surface of the CMB.

In this picture, the CMB anisotropies are ascribed to density fluctuations
originating from quantum fluctuations~\cite{flux} in the scalar field
whose potential
energy $V = {\cal O}(\mu^4)$ drove inflation. These fluctuations would have
induced perturbations in the field energy in different parts of the Universe,
evolving subsequently into fluctuations in the temperature of the CMB. These
would be (approximately) a Gaussian random field of perturbation $\delta\rho /
\rho$, with similar magnitudes on different scale sizes, as favoured by
astrophysicists. The magnitude of these perturbations is
related to the field energy density during inflation
\beq
{\delta T \over T}\sim {\delta\rho\over\rho} \propto \mu^2~G_N
\label{four}
\eeq
The magnitude $\delta T/T\sim 10^{-5}$ observed by COBE et seq. favours
$\mu\lappeq 10^{16}$ GeV, comparable with the unification scale in
GUTs~\cite{GUTs}.
Therefore, at the very least, inflation provides with a unique window
through which we can look
back at an energy scale far beyond the direct reach of current
accelerators, and might even provide us with a precious window on GUTs
themselves. A challenge which has not yet been fully met, however,
is to derive an inflationary potential from some GUT (or string)
theory in a natural way.

\section{What we Might Hope to Learn}

The large mass scale $\mu$ associated with inflation suggests that observations
of the CMB may be sensitive to all mass scales in physics, not excluding that
associated with quantum gravity, which may not be so far beyond $m_{GUT}$, if
current $M$-theory ideas are right.

Indeed, examples can easily be given of the CMB's sensitivity to aspects of
{\bf Quantum~Gravity}. In addition to the scalar density perturbations
(\ref{one}) that are expected to dominate the CMB anisotropies, there may also
be tensor perturbation modes, which are none other than {\it gravitational
waves}. {\it String cosmology} scenarios have been proposed~\cite{Gab}, of
which these may
provide a key observational signature. Then we should recall that it is the
combination of CMB data with those on high-redshift
supernovae~\cite{highz} that provides
the clearest evidence for a {\it cosmological constant} $\Lambda$,
as discussed in more detail later.

The interpretation of these observations corresponds to $\Lambda \lappeq 10^{-123}$ $m^4_P$, which is far
smaller than the individual contributions to $\Lambda$ in many particle
theories. For example, condensates in the QCD vacuum yield
\beq
\delta_{QCD}\Lambda \sim (100~{\rm MeV})^4 \sim 10^{-80} ~m^4_P~,
\label{five}
\eeq
the Higgs vacuum of the Standard Model contributes
\beq
\delta_{EW}\Lambda \sim (100~{\rm MeV})^4 \sim 10^{-68} ~m^4_P~,
\label{six}
\eeq
and global supersymmetry breaking might contribute
\beq
\delta_S \Lambda \gappeq (1~{\rm TeV})^4 \sim 10^{-64} ~m^4_P~,
\label{seven}
\eeq
The discrepancy between these estimates and the (inferred) 
astrophysical value 
may be the biggest problem in particle physics, much bigger even
than the gauge hierarchy problem. Its resolution certainly requires a
consistent quantum theory of gravity that also includes all the other
particle interactions. 

Personally, I regard the observational
indications for non-zero vacuum energy as a tremendous opportunity for theoretical
physics, as it provides a number to calculate in one's candidate theory of
quantum gravity. Much effort has been applied to trying to prove that $\Lambda
= 0$~\cite{whyzero}, but a corresponding exact unbroken symmetry has not
been identified.
Perhaps $\Lambda \not= 0$ after all? Or perhaps it is merely relaxing towards
zero: $\Lambda (t) \rightarrow 0$ with a non-trivial equation of state:
$P/p\equiv \alpha < 0$? Present data require $\alpha \lappeq
-0.6$~\cite{eqstate}, but do not
impose $\alpha = -1$ as required if $\Lambda$ is constant. Models in which
$\Lambda (t)\rightarrow 0$ include a mobile scalar field $\phi(t)$ 
(quintessence) whose potential energy $V(\phi(t))\rightarrow
0$~\cite{quint}, and gradual
de-excitation of the quantum-gravity vacuum~\cite{EMNLambda}. The CMB and
other data may
eventually be able to make interesting distinctions between possible equations
of state, and thereby discriminate between different theories of quantum
gravity.

As for {\bf Grand Unification}, a primary hope is that the vacuum energy
driving inflation could be related to the {\it GUT scalar potential}. 
The CMB may
also cast light on the magnitudes of the {\it neutrino masses} expected in
GUTs. Laboratory experiments have established that these must be much smaller
than the masses $(m)$ of the charged leptons and quarks~\cite{PDG}:
\bea
m_{\nu_e} &\lappeq& 2.5~{\rm eV}, ~~~\\ \nonumber
m_{\nu_\mu} &\lappeq& 160~{\rm keV}, ~~~\\ \nonumber
m_{\nu_\tau} &\lappeq& 18~{\rm MeV}~.
\label{eight}
\eea
Theorists expect non-zero neutrino masses, because there are no candidate exact
gauge symmetries with associated conserved charges to forbid them, by analogy
with the $U(1)_{em}$ of QED, with its associated conserved $Q_{em}$ and 
vanishing photon mass.
We expect the other apparently conserved quantum numbers such baryon number $B$
and lepton number $L$ eventually to be violated, most likely at some high mass
scale $M\simeq M_{GUT}$. Lepton-number violation leads generically to
neutrino masses.

Most theorists expect a see-saw mass matrix mixing the known
$\nu_L$ with heavy singlet neutrinos $N$ 
(often called right-handed neutrinos, but I dislike this
nomenclature) of the form~\cite{seesaw}
\beq
(\nu_L , N)~~\left(\matrix{0 & m \cr m & M}\right)~~\left(\matrix{\nu_L \cr
N}\right)
\label{nine}
\eeq
whose diagonalization yields
\beq
m_\nu \sim {m^2\over M} \ll m \sim m_{l,q}
\label{ten}
\eeq
For example, if we put $m \sim$ 100 GeV and 
take $m_\nu\sim 10^{-1}$ eV for the
third generation, we estimate $M\sim 10^{13}$ GeV. Recent evidence for
atmospheric neutrino oscillations suggests~\cite{SKam}
\beq
\Delta m^2_A \sim (10^{-2}~{\rm to}~ 10^{-3})~{\rm eV}^2
\label{eleven}
\eeq
for the mass-squared difference between one pair of
mass eigenstates $m_{\nu_i}$. The range (\ref{eleven})
can be explored with approved and projected long-baseline
neutrino experiments with accelerator beams~\cite{LBL}. In addition,
solar neutrino data have for some time suggested~\cite{solarnu}
\beq
\Delta m^2_S \sim (10^{-5}~{\rm or}~ 10^{-10})~{\rm eV}^2
\label{twelve}
\eeq
for the mass-squared difference between another pair.
These are not measurements of the absolute scale of neutrino masses, but
most models suggest that the neutrinos are {\it not} heavy and almost
degenerate, and hence that
\bea
m_{\nu_1}&\sim &(10^{-1}~{\rm to}~ 10^{-3/2})~{\rm eV} >\\ \nonumber
 m_{\nu_2} &\sim &
(10^{-5/2}~{\rm or}~ 10^{-5})~{\rm eV} > m_{\nu_3}
\label{thirteen}
\eea
for the three mass eigenstates.
As is discussed below, the CMB and large-scale structure data may eventually
provide the best constraint on the expected hierarchy (13).

 Another
possible output of grand unification that will be constrained by CMB 
measurements is {\it baryogenesis}. Already $\Omega_B$ (and hence $n_B/s$) is
being bounded by present
CMB measurements~\cite{Smoot}, and these may eventually povide the most
accurate
determination of $\Omega_B$, for comparison with baryogenesis scenarios at the
GUT or electroweak scale~\cite{Cline}, or in between.

Another possible extension of the Standard Model that may be tested by CMB
measurements is {\bf Supersymmetry}~\cite{susy}. This is
invoked~\cite{hierarchy} by particle theorists to
stabilize the gauge hierarchy: $m_W \ll m_P$, or equivalently $G_F \sim 1/m^2_W
\gg G_N = 1/m^2_P$, or equivalently $V_{\rm Coulomb} = e^2/r \gg V_{\rm Newton}
= (m_pm_e/m^2_P)~~1/r$ inside an atom. If one tries to set such a hierarchy by
hand, one discovers large quantum corrections:
\beq
\delta m^2_W = {\cal O}\left({\alpha\over\pi}\right)~~\Lambda^2
\label{forteen}
\eeq
which are much larger than the physical value of $m^2_W$ if the cutoff $\Lambda$
in (\ref{forteen})  is ${\cal O}(m_P$ or $m_{GUT}$). An effective cutoff
$\Lambda$ is
provided by sparticle masses in supersymmetric models:
\beq
\delta m^2_W = {\cal O}\left({\alpha\over\pi}\right)~~\vert m^2_B -
m^2_F\vert
\label{fifteen}
\eeq
where the subscripts ($B,F$) denote superpartner bosons and fermions, and
the
remainder is
$\lappeq m^2_W$ if
\beq
\vert m^2_B - m^2_F\vert \lappeq 1~{\rm TeV}^2
\label{sixteen}
\eeq 
This motivates the appearance of superpartners at energies accessible to
accelerators such as the LHC~\cite{LHC}. As discussed below, it also
suggests the
presence of massive supersymmetric relic 
particles contributing ${\cal O}$(1) to the
matter density $\Omega_m$~\cite{EHNOS}. CMB measurements already bound
$\Omega_m$, and may
soon provide accurate measurements of it, thereby constraining supersymmetric
models.

Subsequent epochs of the history of the Universe, such as the {\bf electroweak
phase transition}, the quark-hadron {\bf QCD phase transition} and
cosmological {\bf nucleosynthesis}   will also be constrained by CMB
measurements, but we do not go into details here. 

\section{Density Budget of the Universe}

We phrase our subsequent discussion in terms of  the  density budget of
the Universe, expressed relative to the critical density: $\Omega_i\equiv
\rho_i/\rho_c$.

$\Omega_{tot}$: Inflation suggests that this is practically indistinguishable
from unity: 
$\Omega_{tot} = 1 \pm 0(10^{-4})$, but there are models that predict
$\Omega_{tot} < 1$~\cite{openinflation}. One of these is illustrated in
Fig. 2, which has the potential

\begin{figure}[H]
 \epsfig{figure=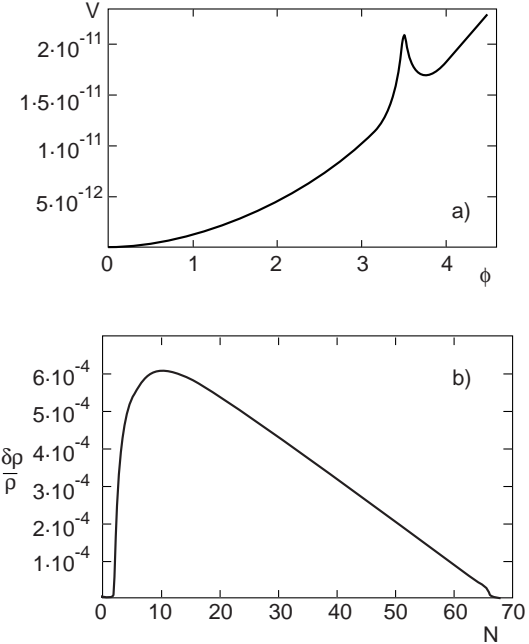,width=7cm}
 \caption[]{{\it a) The potential (17) that leads to open
inflation~\cite{openinflation}, with the spectrum
 b) of density perturbations.}}
\end{figure}

\beq
V = {m^2\phi^2\over 2}~Z~\left(1+{\alpha^2\over
\beta^2+(\phi-v)^2}\right):~\beta\ll v
\label{seventeen}
\eeq
This looks rather bizarre, but who knows what {\it a priori} probability measure
the inflationary God uses, or even whether such a concept makes
sense~\cite{FRW}? As seen in Fig. 2b, this model predicts a spectrum of
density perturbations that is far from flat~\cite{openinflation}, and
hence amenable to test by CMB measurements.

$\Omega_{b}$: Measurements of the $D/H$ ratio in high-redshift Lyman-$\alpha$
clouds~\cite{Tytler} correspond to
\beq
{D\over H} = (3.3 \pm 0.3)\times 10^{-5}
\label{eighteen}
\eeq
If this is indeed the correct primordial $D/H$ ratio, big-bang nucleosynthesis
calculations suggest that~\cite{Turner}
\beq
{n_B\over s} = (5.1 \pm 0.3) \times 10^{-10}
\label{nineteen}
\eeq
corresponding to
\beq
\Omega_B h^2 = 0.019 \pm 0.001
\label{twenty}
\eeq
where $h$ is the present Hubble expansion rate in units of 100 kms$^{-1}
M_{\rm pc}^{\phantom{\rm pc}-1}$.
Using the currently favoured range $h = 0.65 \pm 0.10$, we
see from
(\ref{twenty}) that $\Omega_b \lappeq 0.08$, which is insufficient to explain
all the matter density in the following paragraph.

$\Omega_m$: The cluster measurements ($M/L$ ratio, present and past abundances,
cluster dynamics and the baryon fractions inferred from $X$-ray measurements)
all suggest~\cite{Bahcall}
\beq
\Omega_m \sim 0.2~~{\rm to}~~ 0.3
\label{twentyone}
\eeq
Moreover,  the combination of CMB measurements and
high-redshift supernovae~\cite{highz} also support 
independently such a value for $\Omega_m$.

$\Omega_{CDM}$: The theory of large-scale structure formation strongly suggests
that most of $\Omega_m$ is cold dark matter, so that
\beq
\Omega_{CDM} \sim \Omega_m
\label{twentytwo}
\eeq
as perhaps provided by supersymmetric particles.

The lightest supersymmetric particle is expected to be stable in most models,
and hence present in the Universe today as a cosmological relic from the Big
Bang~\cite{EHNOS}. Its stability would be due to a
multiplicatively-conserved quantum
number, called $R$ parity, which is related to baryon number $B$, lepton number
$L$ and spin $S$:
\beq
B = (-1)^{3B+L+2S}
\label{twentythree}
\eeq
and takes the value +1 for all conventional particles, changing to -1 for all
supersymmetric particles, because they have identical internal properties but
spins differing by half a unit. There are three important consequences of $R$
conservation: (i) sparticles  are always produced in pairs, such as $\bar pp
\rightarrow \tilde q \tilde g + X$ or $e^+e^- \rightarrow
\tilde\mu^+\tilde\mu^-$, (ii) heavier sparticles decay into lighter ones, such
as $\tilde q \rightarrow q \tilde g$ or $\tilde\mu \rightarrow \mu\tilde\gamma$,
and (iii) the lightest sparticle is stable because it has no legal decay mode.

In many models~\cite{EHNOS}, the favoured scandidate for the lightest
sparticle is the lightest neutralino
$\chi$, which is a mixture of the photino $\tilde\gamma$, the zino $\tilde Z$ and
the neutral Higgsinos $\tilde H^0$~\cite{Xgluino}. At the tree level, the
neutralinos are
characterized by three parameters: the unmixed gaugino mass $m_{1/2}$, a Higgsino
mixing parameter $\mu$ and $\tan\beta$, the ratio of Higgs vacuum expectation
values. The properties of the $\chi$ particle simplify in the limit
$m_{1/2}\rightarrow 0$, where it becomes an almost pure photino $\tilde\gamma$,
and in the limit
$\mu\rightarrow 0$, where it becomes almost a pure Higgsino $\tilde H$.
However, the non-observation of supersymmetric particles at LEP excludes these
simple limits~\cite{EFOS}. The purely experimental limit $m_\chi \gappeq$
20 to 30 GeV may be strengthened by taking other
constraints into account~\cite{EFOS,EFGOS}, as seen in Fig. 3.

There are generic domains of supersymmetric parameter space where an
``interesting"  cosmological relic density $0.1 \lappeq \Omega_\chi h^2
\lappeq 0.3$ is possible~\cite{EHNOS} and it can even 
be argued that this is the most natural
range~\cite{CEOPO}. If this upper limit is imposed, the
lower bound on
$m_\chi$ is strengthened to the dotted line marked $C$ in Fig. 3. The limit
coming from the non-observation of a supersymmetric Higgs boson at LEP is
indicated by the dotted line marked $H$ in Fig. 3, which is strengthened to
the solid line marked UHM if all the scalar sparticles are assumed to have the
same mass as the Higgs fields at the GUT input scale. Finally, combining this
assumption with the lower and upper limits on the cosmological relic density
yields the lines marked DM + UHM and cosmo + UHM in Fig. 3. These
considerations currently yield~\cite{EFGOS}
\beq
m_\chi \gappeq 42~{\rm GeV}
\label{twentyfour}
\eeq
and subsequent LEP runs should be able to explore thoroughly the range $m_\chi
\lappeq$ 50 GeV.

Although theorists of structure formation prefer most of the dark matter to be
composed of cold non-relativistic particles, such as neutralinos, they think
this may not be the whole story, as seen in Fig. 4~\cite{Turnerrev}. The
plain CDM model would
require a very non-flat spectrum of perturbation $n \ll 1$, which is disfavoured
in most inflationary models, if $h
\simeq 0.65$, as suggested by
current data. A model
($\tau$CDM) with decaying dark matter fares somewhat better, but the most
promising are the mixed ($\nu$CDM) model and the model ($\Lambda$CDM) with a
cosmological constant.

$\Omega_{HDM}$: The hot dark matter density due to neutrinos can be predicted
accurately as a function of the neutrino masses
\beq
\Omega_{HDM} h^2 \sim \sum_i~~\left({m_{\nu_i}\over 98~{\rm eV}}\right)
\label{twentyfive}
\eeq
The theory of structure formation suggests that $\Omega_{HDM} \ll
\Omega_{CDM}$, and the indications (\ref{eleven}), (\ref{twelve}) from
atmospheric and solar neutrino data can easily be explained (\ref{thirteen}) by
light neutrinos: $m_{\nu_i} <$ 0.1 eV, which would make a  small
contribution to $\Omega_{tot}$.

\begin{figure}
\epsfig{figure=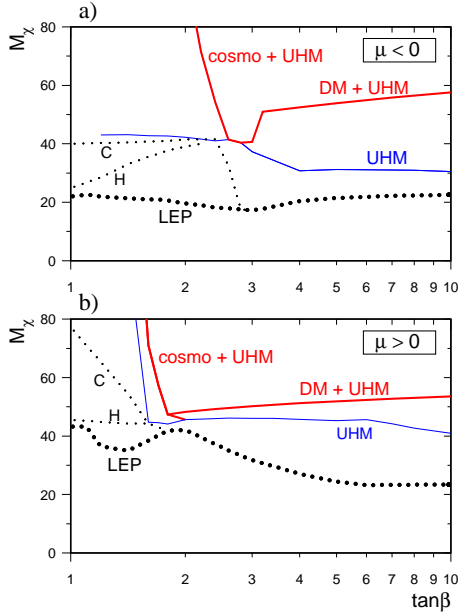,width=7cm}
 \caption[]{{\it Lower limits on the lightest neutralino
mass~\cite{EFGOS},
obtained under the different assumptions listed in the text.}}
\end{figure}

\begin{figure}
 \epsfig{figure=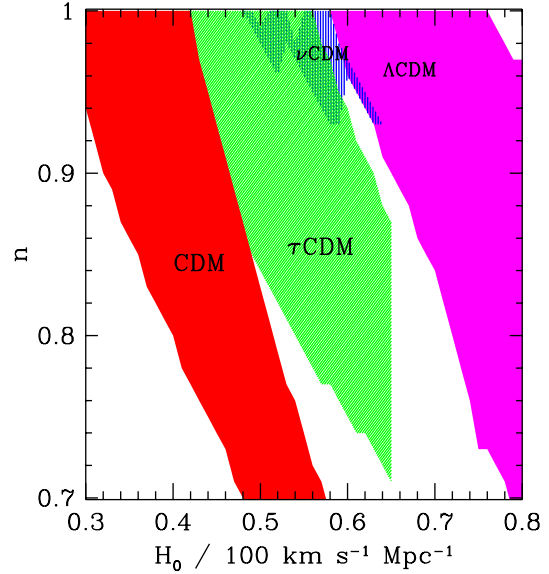,width=7cm}
 \caption[]{{\it Allowed ranges~\cite{Turnerrev} of the Hubble expansion
rate and the power law for
 cosmological perturbations, in different dark matter models.}}
\end{figure}

The present and prospective sensitivities of cosmological data to $m_\nu$ are
shown in Fig. 5~\cite{Tegmark}. So far, $m_\nu \gappeq$ 3 eV is excluded
by the available
upper limit on the density of hot dark matter. Comparison of future data on
large-scale structure and the CMB are thought to be sensitive to $m_\nu
\gappeq$ 0.3 eV. This is very close to the range $m_\nu \sim$ 0.1 to 0.03 eV
favoured by the atmospheric neutrino data, so one should not abandon hope of
detecting neutrino masses astrophysically~\cite{betabeta}.

$\Omega_\Lambda$: As we have already seen, the largest fraction of the energy
density of the Universe may be provided by vacuum energy, if one combines the
CMB~\cite{anisotropies} and high-redshift supernova data~\cite{highz}. It
is also required by the dynamical
estimates of $\Omega_m$ and inflation, which requires $\Omega_{tot} = \Omega_m
+ \Omega_\Lambda \simeq 1$.

If one takes at face value the absolute scale of neutrino masses
suggested by the atmospheric neutrino data, one would be
led to favour the ($\Lambda$CDM) option in Fig. 4. In this case, Fig. 6
pieces together the indications concerning $\Omega_\Lambda$ and $H_0$ from
different astrophysical and cosmological data {\it excluding} those on
high-redshift supernovae. We see that these favour {\it independently}
$\Omega_\Lambda \sim$ 0.6, $h \sim 0.65$~\cite{Turnerrev}. Thus a
remarkably consistent picture of the density budget of the Universe may be
emerging:

\begin{figure}[H]
 \epsfig{figure=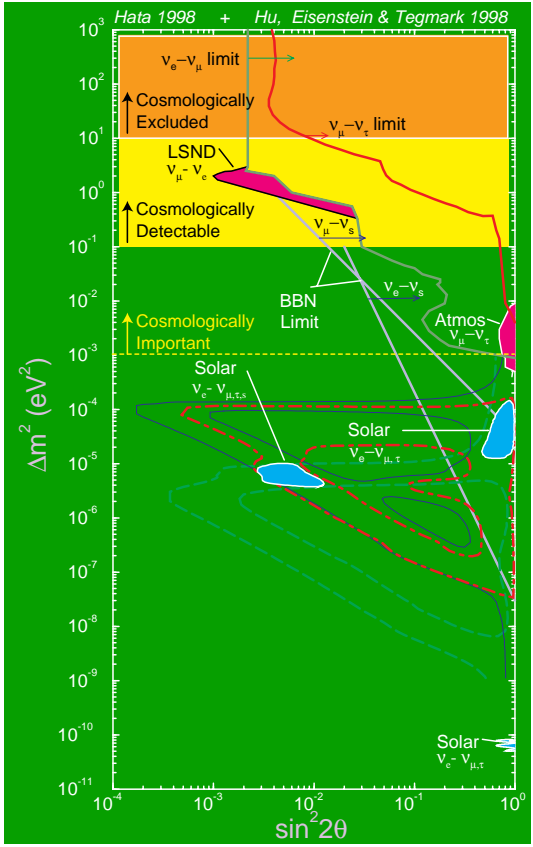,width=7cm}
 \caption[]{{\it Compilation of indications on neutrino mass-squared
differences $
 \Delta m^2$ and mixing angles $\theta$ from oscillation experiments, compared
 with cosmological sensitivities to neutrino masses~\cite{Tegmark}.}}
\end{figure}

\beq
\Omega_{tot} \simeq 1 = \Omega_m + \Omega_\Lambda : \\
\Omega_m \sim 0.3, \Omega_\Lambda \sim 0.7
\label{twentysix}
\eeq
where
\beq
\Omega_m = \Omega_{CDM} + \Omega_\nu + \Omega_b
\label{twentyseven}
\eeq
with
\beq
\Omega_b < 0.1~,~~~\Omega_\nu \ll \Omega_{CDM} \simeq \Omega_m
\label{twentyeight}
\eeq
Let us see whether future data confirm this picture.

\section{What we Have Learnt}

The first generation of CMB measurements has already taught us a great deal
about fundamental physics~\cite{Smoot}, some of which has already been
mentioned in 
previous sections. Most of the discussion is in terms of inflationary models,
but it should not be forgotten that the CMB delivered a death blow to the
alternative models based on cosmological defects~\cite{Turok}. These did
not predict
\begin{figure}
 \epsfig{figure=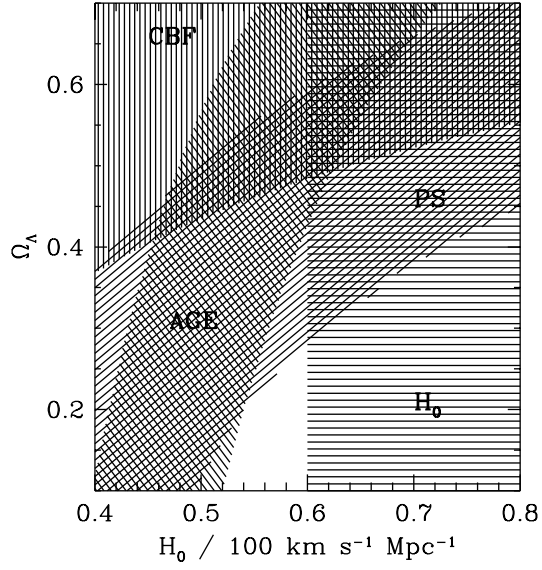,width=7cm}
 \caption[]{{\it Convergent indications of a non-zero cosmological
constant~\cite{Turnerrev}, independent of high-$z$ supernovae, 
from large-scale structure and CMB anisotropies (PS), the age of
the Universe (AGE), the fraction of baryons in clusters (CBF) and
measurements of $H_0$.}}
\end{figure}
\noindent
an
acoustic peak, as apparently observed in the data at a harmonic number
$\ell\sim$ 200, as suggested by the data compiled in Fig.~1. 
It is mainly the location of this peak that suggests $\Omega_{tot}=
\Omega_m + \Omega_\Lambda \simeq 1$, as seen in Fig. 7. The height of the peak,
as seen in Fig. 1,
suggests that $\Omega_b \lappeq 0.1$. Moreover, the
combination of CMB with large-scale structure suggests that $\Omega_{CDM} \gg
\Omega_{HDM}$, and the value of $\Omega_{CDM}$ suggested by combining the CMB
data with high-redshift supernovae is compatible with $\Omega_{CDM} \sim 0.3$
(as also seen in Fig.~7~\cite{combine}) as suggested by cluster
observations. As has already
been mentioned, standard cold dark matter does not fit the CMB and large-scale
structure data, but a model with $\Omega_\Lambda \sim 0.7$ does. 
Furthermore, the indications from the CMB and large-scale structure data are
that the spectral index of the density perturbations $n \sim 1 \pm 0.2$, in
agreement with the Harrison-Zeldovich spectrum and most
inflationary models.

\begin{figure}
 \epsfig{figure=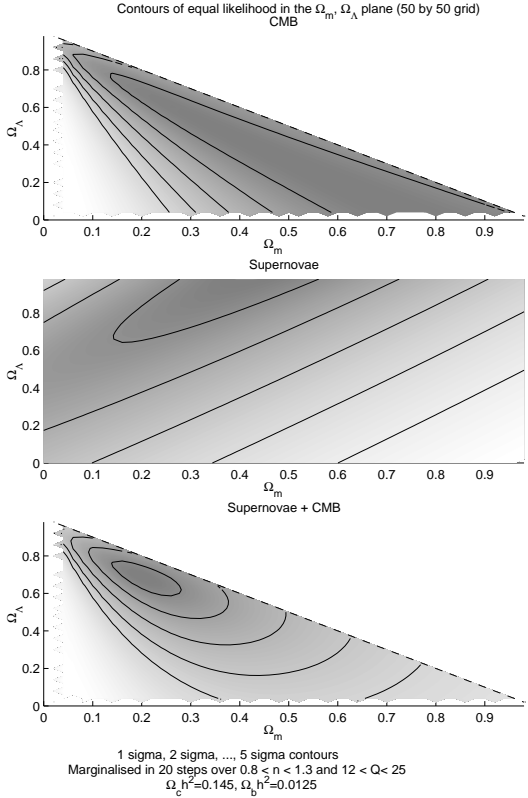,width=7cm}
\caption[]{{\it Indications on $\Omega_m$ and $\Omega_\Lambda$ from the
combination of
CMB fluctuations and high-$z$ supernovae~\cite{combine}.}}
\end{figure}

Thus, we have reached the stage where alternative theories of structure
formation have been, or are being, eliminated, and attention is being focussed
on a candidate Standard Model of structure formation. The next step is test the
model, determine its parameters, and try to over-constrain them, with the hope
of eventually moving beyond it.

\section{What we May Hope to Learn}

The next generation of CMB measurements, culminating in the MAP and Planck
satellites, will provide us with precision determinations of physical
quantities, and probe the emerging Standard Model of structure formation, much as
LEP and the SLC have probed the Standard Model of particle physics. For
example, $\Omega_{tot}$ may be determined with a precision of 0.1, possibly
0.01 in combination with high-redshift supernova data, in conjunction with a
comparable precision in $\Omega_m$. Similarly, $\Omega_b$ will be determined
with a small fractional error. In the case of LEP, many quantities such as
$m_Z$, $\sin^2\theta_W$ and the number of light neutrino species $N_\nu$ 
were eventually determined with errors far smaller than theoretical guesses
before the accelerator started. For that reason, I am not going to hazard here
many guesses about the eventual errors in cosmological parameters! However, let us
consider neutrinos as an example of what may be possible.

These decoupled when the temperature was ${\cal O}(1)$ MeV. Following
reheating by
$e^+e^-\rightarrow \bar\nu\nu$, we expect a relic density
\beq
{\rho_\nu\over\rho_\gamma} = {7\over 8}~~\left({4\over 11}\right) ^{4/3} N_\nu
\simeq 0.681 N_\nu
\label{twentynine}
\eeq
This is subject to small corrections due to incomplete decoupling: $\delta
N_\nu^{ID} \simeq$ 0.03 to 0.04 and finite-temperature QED corrections:
$\delta N^{FT}_\nu \simeq$ 0.01~\cite{nucount}. The precise value of the
ratio $\rho_\nu
/\rho_\gamma$ affects the epoch of matter-radiation equality, and can be measured
accurately by the Planck satellite, particularly using
polarization~\cite{nucount}.

Figure 8 
shows the predicted sensitivity to $\delta N_\nu$ as a function of the maximum
value of $\ell$, demonstrating the advantages gained from polarization data and
from measurements at high $\ell$. These may reach the sensitivity required to
see the effects of incomplete decoupling and finite-temperature
QED~\cite{nucount}. They may
even match the LEP error $N_\nu^{LEP} = 2.994 \pm 0.011$~\cite{LEPEWWG}!

\begin{figure}
 \epsfig{figure=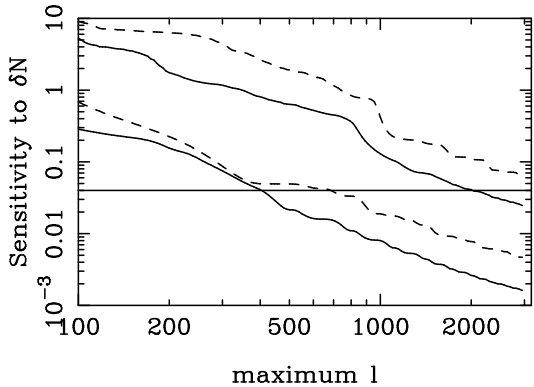,width=7cm}
 \caption[]{{\it Possible sensitivity of future CMB measurements to the
effective
 number of neutrino species~\cite{nucount}. The lower (upper) pair of
lines (do not) assume exact knowledge of other cosmological
parameters. Within each pair, the lower (solid)
line is the sensitivity obtainable if polarization is measured.}}
\end{figure}

CMB measurements, in conjunction with other observations,
will also be able to put interesting constraints on the mass of any stable
neutrino, in the range of 1 to possibly 0.1 eV~\cite{Tegmark}. As
already mentioned, this is getting close to the
range indicated by atmospheric neutrino oscillations:
\beq
m_\nu \sim \sqrt{\Delta m^2_{Atmo}} \sim 0.1~~{\rm to}~~ 0.03~~{\rm eV}
\label{thirty}
\eeq
as seen in Fig. 5. 
I would not bet a lot of money against the CMB and large-scale structure data
eventually reaching down to the range (\ref{thirty}). The present CMB data are
already able to exclude decaying neutrinos with $m_\nu \gappeq$ 10 eV and
$10^{13} s \lappeq \tau \lappeq 10^{17} s$~\cite{nudecay}. Again, the
future sensitivity is
expected to extend down to about 1 eV, and there will be analogous constraints
on other unstable massive particles such as neutralinos, gravitinos, etc.

The future CMB measurements will also make precision tests of inflationary
models, much as LEP and the SLC have made precision tests of electroweak models
and measured $\sin^2\theta_W$ very accurately. The observables of interest are
the scalar perturbation mode $S$, the tensor mode $T$, and their spectral
indices $n, n_T$. Knowledge of them enables the inflationary potential to be
reconstructed~\cite{reconstruction}:
\bea
V_* &\simeq &1.65 T~m^4_P~,~~\\ \nonumber
 V^\prime_* &\simeq& ~\pm\sqrt{{8\pi\over 7}~{T\over
S}}~~{V_*\over m_P}~,\\ \nonumber
V^{\prime\prime}_* &=& 4\pi \left[(n-1)+{3\over 7}~{T\over
S}\right] ~~{V_*\over m^2_P}
\label{thirtyone}
\eea
where the primes denote derivatives with respect to the inflaton field $\phi$,
and the $\ast$ subscript  denotes the scale at which the measurement is made. In
addition, there is a consistency condition
\beq
{T\over S} = -7 n_T
\label{thirtytwo}
\eeq
which enables the inflationary paradigm to be checked. Figure 9 shows how the
spectral index $n$ and the tensor/scalar ratio $r$ vary in different inflationary
models~\cite{Gain}. Also shown are the error ellipses expected from
Planck. We see
that the latter should be able to distinguish between different power-law
potential models, and between many of these and models with an exponential
potential.

\begin{figure}
 \epsfig{figure=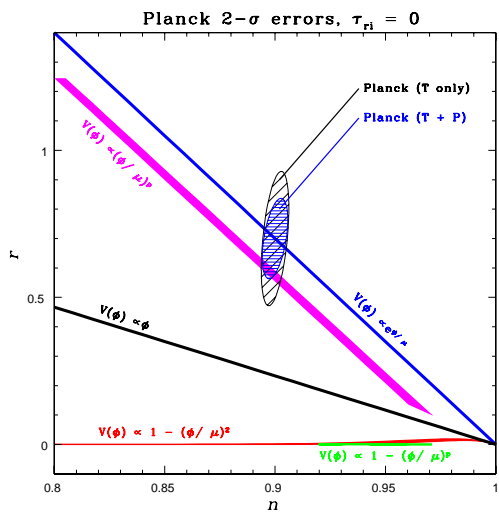,width=7cm}
 \caption[]{{\it Likely future precision in CMB measurements confronted
with various
 model predictions~\cite{Gain}.}}
\end{figure}
 
CMB measurements take inflationary models into the scientifc
domain: individual models may be falsified, and even very general classes
of models, for example by observing strong non-Gaussian correlations.
A word of caution is, however, in order. Like any finite-size set of
measurements, the CMB measurements alone will not have a unique interpretation
-- the so-called cosmological degeneracy problem. Specifically, models with
the same stress history will give the same pattern of acoustic peaks in the
CMB spectrum and the same perturbation power  spectrum. Other measurements
sensitive to the co-moving sound speed and curvature fluctuations would be
needed to distinguish between models.

\section{Conclusions}

The CMB is a powerful probe of fundamental physics, including quantum gravity,
inflation, grand unification, cold dark matter, hot dark matter, decaying
particles, Big-Bang baryosynthesis and much else besides, as well as being of
capital importance for astrophysics and cosmology. It is one of the very few
probes we have of physics at the grand unification scale, along with neutrino
physics, as can be probed using long-baseline neutrino
experiments~\cite{LBL},
and the measurements of gauge couplings and particle masses, e.g., of
sparticles at the LHC~\cite{LHC}. Hence the future generation of CMB
experiments, notably
MAP and particularly Planck, is an invaluable  complement to the next
generation of particle accelerator experiments at the LHC. Together, they may
not only establish a Standard Model of cosmology and structure formation, but
also take us beyond the Standard Model of particle physics.


\begin{thebibliography}{99}

\bibitem{Smoot} G.F. Smoot, astro-ph/9902027.

\bibitem{PDG} Particle Data Group, C. Caso et al.,
Eur.Phys.J., C3 (1998) 1.

\bibitem{FRW} For a review with a constructively critical attitude,
see:\\
G.F.R. Ellis, gr-qc/9812046.

\bibitem{dipole} A great scientific check of this interpretation is
to compute successfully the dipole anisotropy using knowledge of the local
distribution of matter. see, e.g., \\
M. Hudson, A. Dekel, S. Courteau, S. Faber and J.A. Willick,
Mon.Not.Roy.Astron.Soc., 274 (1995) 305.

\bibitem{anisotropies} for a recent compilation, see: M. Tegmark,
{\tt http://www.sns.ias.edu/${\sim}$max/r{\_}frames.\\html}.


\bibitem{inflation} A.A. Starobinsky, \PL 91B (1980) 99;\\
D. Kazanas, \AJ  241 (1980) L59;\\
A. Guth, \PR  D23 (1981) 347.

\bibitem{monopoles} J. Preskill, \PRL  43 (1979) 1365;\\
Ya. Zel'dovich and M. Khlopov, \PL  79B (1979) 239.

\bibitem{flux} J. Bardeen, P.J. Steinhardt and M.S. Turner,
\PR D28 (1983) 679;\\
A.H. Guth and S.-Y. Pi, \PRL 49 (1982) 1110;\\
A.A. Starobinsky, \PL 117B (1982) 175;\\
S.W. Hawking, \PL 115B (1982) 295

\bibitem{GUTs} J. Ellis, S. Kelley and D.V. Nanopoulos, \PL  B249
(1990) 441 and \PL  B260 (1991) 131;\\
U. Amaldi, W. de Boer and H. Furstenau, \PL  B260 (1991) 447;\\
P. Langacker and M. Luo, \PR  D44 (1991) 817.

\bibitem{Gab} G. Veneziano, hep-th/9902097 and references therein.

\bibitem{highz} A.G. Riess et al., astro-ph/9805201; \\
S. Perlmutter et al., astro-ph/9812133.

\bibitem{whyzero} S. Kachru and E. Silverstein, hep-th/9810129.

\bibitem{eqstate} P.M. Garnavich et al., astro-ph/9806396; \\
S. Perlmutter, M.S. Turner and M. White, astro-ph/9901052.

\bibitem{quint} I. Zlatev, L.-M. Wang and P.J. Steinhardt,
astro-ph/9807002; \\
P.J. Steinhardt, L.-M. Wang and I. Zlatev, astro-ph/9812313.

\bibitem{EMNLambda} J. Ellis, N. Mavromatos and D. Nanopoulos,
gr-qc/9810086.

\bibitem{seesaw} T. Yanagida, Proc. Workshop on the
Unified Theory and the Baryon Number in the Universe (KEK, Japan, 1979);\\
R. Slansky, Talk at the Sanibel Symposium, Caltech preprint CALT-68-709
(1979).

\bibitem{SKam} Super-Kamiokande collaboration, Y. Fukuda et al.,
\PRL 81 (1998) 1562.

\bibitem{LBL} Y. Oyama, for the K2K collaboration, hep-ex/9803014;\\
MINOS collaboration, E. Ables et al., Fermilab proposal  P-875 (1995);\\
G. Acquistapace et al., CERN report 98-02 (1998).

\bibitem{solarnu} J.N. Bahcall, astro-ph/9808162.

\bibitem{Cline} J.M. Cline, hep-ph/9902328.

\bibitem{susy}  P. Fayet and S. Ferrara, {Phys.Rep.,}  32,
251 (1977); \\
H.E. Haber and G.L. Kane, {Phys.Rep.,}  117,  75 (1985).

\bibitem{hierarchy} 
L. Maiani, {\it Proc. Summer School on Particle
Physics},
Gif-sur-Yvette, 1979 (IN2P3, Paris, 1980) p. 3;\\
G 't Hooft, in:  G 't Hooft et al., eds., {\it Recent Developments in
Field
Theories}
(Plenum Press, New York, 1980);\\
E. Witten, \NP  B188  513 (1981);\\
R.K. Kaul, \PL  109B  19 (1982).

\bibitem{LHC} S. Abdullin and F. Charles, hep-ph/9811402.

\bibitem{EHNOS} J. Ellis, J.S. Hagelin, D.V. Nanopoulos, K.A. Olive
and M. Srednicki, \NP  B238, 453 (1984).

\bibitem{openinflation} A. Linde, \PR D59 (1999) 023503.

\bibitem{Tytler} D. Tytler, S. Burles, L.-M. Wu, X.-M. Fan, A. Wolfe and
B.D. Savage, astro-ph/9810217.

\bibitem{Turner} S. Burles, K.M. Nollett, J.N. Truran and M.S. Turner,
astro-ph/9901157.

\bibitem{Bahcall} N.A. Bahcall, astro-ph/9901076.

\bibitem{Xgluino} Any charged or strongly-interacting sparticle,
including the gluino, is probably excluded as a cosmological relic
by the stringent experimental upper limits on the abundances of anomalous
heavy isotopes~\cite{EHNOS}.

\bibitem{EFOS} J. Ellis, T. Falk, K. Olive and M. Schmitt, \PL {B388}
(1996) 97 and \PL  B413, 355 (1997).

\bibitem{EFGOS} J. Ellis, T. Falk, G. Ganis, K.A. Olive and M. Schmitt,
\PR D58 (1998) 095002.

\bibitem{CEOPO} P.H. Chankowski, J. Ellis, K.A. Olive and S. Pokorski,
hep-ph/9811284.

\bibitem{Turnerrev} S. Dodelson, E.I. Gates and M.S. Turner,
Science, 274 (1996) 69; \\
M.S. Turner, astro-ph/9901168.

\bibitem{Tegmark} W. Hu, D.J. Eisenstein and M. Tegmark,
\PRL 80 (1998) 5255 and \\
{\tt http://www.sns.ias.edu/{$\sim$}whu/pub.html}.

\bibitem{betabeta} Searches for neutrinoless double-$\beta$ decay
are sensitive to $<m_{\nu}>_e \sim 0.2$~eV, where the 
neutrino masses are weighted by their couplings to electrons:
L. Baudis et al., hep-ex/9902014.

\bibitem{Turok} U.-L. Pen, U. Seljak and N.G. Turok, \PRL 79
(1997) 1611.

\bibitem{combine} A.N. Lasenby, S.L. Bridle and M.P. Hobson,
astro-ph/9901303.

\bibitem{nucount} R.E. Lopez, S. Dodelson, A. Heckler and M.S. Turner,
astro-ph/9803095.

\bibitem{LEPEWWG}  M. Gr\"unewald and D. Karlen, talks at International
Conference on High-Energy Physics, Vancouver 1998, \\
{\tt http://www.cern.ch/LEPEWWG/misc}.

\bibitem{nudecay} R.E. Lopez, S. Dodelson, R.J. Scherrer and M.S. Turner,
\PRL 81 (1998) 3075, and references therein.

\bibitem{reconstruction} E.J. Copeland, I.J. Grivell, E.W. Kolb and A.R.
Liddle, \PR D58 (1998) 043002, and references therein.

\bibitem{Gain} W.H. Kinney, \PR D58 (1998) 12350.

\end{thebibliography}
\end{document}